\documentclass{article}

\usepackage[english]{babel}
\usepackage[letterpaper,top=0.74in,bottom=2cm,left=0.5in,right=0.5in,marginparwidth=0.5in]{geometry}

\makeatletter
\if@titlepage
  \renewenvironment{abstract}{%
      \titlepage
      \null\vfil
      \@beginparpenalty\@lowpenalty
        \paragraph{\abstractname:}
        \@endparpenalty\@M
      }%
     {\par\vfil\null\endtitlepage}
\else
  \renewenvironment{abstract}{%
      \if@twocolumn
        \section*{\abstractname}%
      \else
        \small
        \paragraph{\abstractname:}
      \fi}
      {\if@twocolumn\else\par\bigskip\fi}
\fi
\makeatother

\usepackage{amsmath}
\usepackage{amssymb}
\usepackage{graphicx}
\usepackage[colorlinks=true, allcolors=blue]{hyperref}

\title{Negative Shaping Order $K$ in Set Shaping Theory: A Comprehensive Analysis}
\author{Sochima Biereagu\footnote{Author: sochima.eb@gmail.com}}
\date{}

\begin{document}
\maketitle

\abstract{
This paper explores an innovative aspect of the Set Shaping Theory \cite{set_shaping_theory}, the use of a negative shaping order $K$. Traditionally, the theory utilizes a positive $K$ \cite{shaping_order_intro} to extend the length of data strings, enhancing their testability \cite{testability_kozlov} and compressibility \cite{shannon}. We propose a paradigm shift by employing a negative $K$, which shortens data strings and potentially improves compression efficiency. However, this approach sacrifices the local testability of the data, a cornerstone in traditional Set Shaping Theory. We examine the theoretical implications, practical benefits, and challenges of this new methodology.
}

\section*{Introduction}
The Set Shaping Theory \cite{set_shaping_theory}, since its inception, has revolved around the utilization of a positive shaping order $K$ to transform data strings. However, recent developments suggest that employing a negative $K$ can significantly optimize compression, especially when coupling high entropy sequences with longer strings. This innovative approach potentially marks a paradigm shift in data compression methodologies, offering a deeper understanding of the $K$ parameter and its implications.

\subsubsection*{Brief Overview of Set Shaping Theory}
Fundamentally, the Set Shaping Theory investigates bijection functions that transform one set of strings into another. The crux is the transformation of string sets, such as $X^N$, into longer string sets, like $Y^{N+K}$, with each string in the latter being $K$ symbols longer than its counterpart in the former.

\subsubsection*{Definitions and Basic Concepts}
\begin{itemize}
    \item \textbf{Source Representation}: A source, denoted as $X$, is characterized by the triplet $X = (x; A; P)$, where:
    \begin{itemize}
        \item $x$ is the value of a random variable.
        \item $A$ represents the possible states or values of $x$.
        \item $P$ is the probability distribution associated with the states.
        \item We call $X^N$ the set of all possible strings $x = x_1, \ldots, x_j, \ldots, x_N$ generated by $X$.
    \end{itemize}
    \item \textbf{Entropy \cite{shannon}}: A measure of unpredictability or randomness of a set, $H(X)$, is given by: 
    $$H(X) = -\sum_i p_i \log_b p_i$$

    \item \textbf{Information Content (zero-order empirical entropy)}: is a measure of uncertainty or randomness associated with a discrete probability distribution, For a specific sequence $x$ in $X^N$, its information content, $I(x)$, is:
    $$I(x) = -\sum_{j=1}^N \log_2 p(x_j)$$
    where $p(x_j)$ refers to the probability of the occurrence of the symbol $x_j$ in the sequence $x$.
    \newline\newline\textbf{Note: In this article, the information content always refers to the zero-order empirical entropy.}
\end{itemize}

\subsubsection*{}
The bijection function $f$ that performs the transform is defined as:
$$f: X^N \rightarrow Y^{N+K}$$

The shaping order, $K$ \cite{shaping_order_intro} acts as a determinant of how much length is added to each string during the transformation. If you're transforming a string from the set $X^N$ using the function $f$, the resulting string in $Y^{N+K}$ will be $K$ characters/symbols longer.
\newline\newline
\textbf{Transformation process:}
\begin{itemize}
    \item Using the bijection function $f$, a string $x$ in $X^N$ is mapped to a new string $y$ in $Y^{N+K}$
    \item The key insight from set shaping theory is that the strings in $Y^{N+K}$ are chosen based on their lower information content from the set $X^{N+K}$. This means that, even though the strings in $Y^{N+K}$ are longer, they might be more "compressible" due to their reduced information content.
\end{itemize}

\subsection*{Local testability}

Local testability \cite{goldreich2017} \cite{cheraghchi2005} is a fundamental concept in data integrity and error detection, playing a pivotal role in modern data processing and communication systems. It refers to the ability to efficiently test the integrity of a piece of data without the necessity to examine the entire data set. This capability is crucial in large-scale data environments where scanning entire datasets for errors is impractical.
\subsubsection*{Set Shaping Theory and Local Testability}
In Set Shaping Theory, local testability stands out as a significant feature due to its efficient approach to maintaining data integrity \cite{testability_kozlov}. In traditional error correction techniques, redundancy is incorporated within the codewords, Set Shaping Theory offers an alternate method. The essence of this method revolves around the bijection functions $f(X^N) = Y^{N+K}$ that transform one set of strings into another of equal size but with strings of greater length. The transformation process conditions the emission probability of the dependent variable $y$ based on previously emitted variables. As a consequence, if a decoder encounters a symbol with a conditional probability of zero, it is an immediate indication of an error in the message, facilitating local testability \cite{testability_kozlov}.
\newline\newline
Several advantages arise from this Approach:
\begin{itemize}
    \item \textbf{Efficient Local Testing:} The reshaped strings in $Y^{N+K}$ are selected for their minimized average information content, which streamlines the process of local testing. This efficiency arises from the predictability and structured nature of these strings, making anomalies more detectable.
    \item \textbf{Enhanced Error Detection through Reduced Entropy:} The selection of strings based on lower entropy not only facilitates better compression but also contributes to more effective error detection. Lower entropy in strings means fewer variations and unexpected elements, allowing for easier identification of errors during local testing.
    \item \textbf{Intrinsic Error Detection Capability:} The unique feature of the Set Shaping Theory is its intrinsic error detection mechanism. By conditioning the emission probabilities of each symbol in the string, the theory enables the immediate identification of errors. A symbol with a conditional probability of zero is a clear indicator of a fault in the message, thus providing a robust method for error detection without the need for extensive data scanning.
\end{itemize}

\subsection*{Rationale for exploring a negative $K$ in Set Shaping Theory.}

Set Shaping Theory, originally designed for transforming and compressing data with positive shaping orders, mainly extends data strings. This usual method using a positive $K$ improves compressibility and keeps local testability. Yet, always using positive $K$ values might miss out on more efficient data compression. This has sparked interest in studying the effects of a negative $K$ in Set Shaping Theory.
\newline\newline
Some reasons to explore Negative $K$:
\begin{itemize}
    \item \textbf{Pushing the Boundaries of Compression:} The primary motivation behind considering a negative $K$ is the pursuit of higher compression efficiency. By shortening data strings, there is an inherent potential for reducing the overall data footprint significantly.
    \item \textbf{Theoretical Novelty and Exploration:} From a theoretical standpoint, extending the Set Shaping Theory to include negative $K$ values opens new avenues of research. It challenges the conventional wisdom of data transformation and presents an opportunity to deepen our understanding of the interplay between string length, information content, and compressibility.
    \item \textbf{Practical Implications in High-Entropy Contexts:} In practical applications involving high-entropy data sets, a negative $K$ could yield considerable advantages. These contexts often involve managing large volumes of data where conventional compression methods may fall short. Employing a negative $K$ could potentially unlock new methods of handling such data more efficiently.
    \item \textbf{Exploring Trade-offs:} The exploration of a negative $K$ is not without its trade-offs, primarily the loss of local testability. This exploration is crucial for understanding the limits and potential drawbacks of the Set Shaping Theory when applied in unconventional ways. It allows for a comprehensive understanding of the theory's applicability and limitations.
\end{itemize}

\section*{Theoretical Foundation of Negative $K$ }

Set Shaping Theory traditionally explores bijection functions $f: X^N \rightarrow Y^{N+K}$ with a single shaping order $K$, where $N, K \in \mathbb{N}^+$, and $|X^N| = |Y^{N+K}|$. This implies a direct one-to-one correspondence between the elements of the sets, preserving their cardinality. Additionally, the subset relationship $Y^{N+K} \subset X^{N+K}$ holds, indicating that the transformed set is a part of the extended original set.
\newline\newline
When considering a negative shaping order $K$, the dynamics of this bijection undergo a significant alteration. With $K$ being negative, the length of strings in the transformed set $Y$ is reduced compared to their counterparts in set $X$. This reduction in length introduces challenges in maintaining the bijection, primarily due to the potential decrease in the total number of unique strings that can be formed with a shorter length.

To circumvent this challenge and preserve the bijection when employing negative $K$ values in Set Shaping Theory, a nuanced strategy that involves the integration of both negative and positive $K$ values is essential. This approach ensures that the transformation maintains a one-to-one correspondence and effectively utilizes residual sequences, which are pivotal in this context

\subsubsection*{Integrating Negative $K$ with Positive $K$}

In the application of negative $K$ values, it is crucial to first apply only positive $K$ values for the initial $|-K|$ values of $N$. This method ensures that the zero-order empirical entropy is increased appropriately, setting a foundation for the subsequent application of negative $K$. For instance, in a scenario where $K = -6$, the transformation with positive $K$ is applied for the first six values of $N$ ($1$, $2$, $3$, $4$, $5$, $6$). Once $N$ exceeds this range, negative $K$ values are employed, utilizing the residual sequences left from the transformations at smaller values of $N$.

\subsubsection*{Example of Applying Negative $K$}
To illustrate this concept further, let's examine a binary example ($A=\{0,1\}$) where $N=3$ and $K=1, -2$:

\begin{itemize}
    \item For $N = 1$, we can only apply positive $K$. The transformation is as follows:
\begin{itemize}
    \item 0 becomes 00
    \item 1 becomes 11
    \item Residual sequences of length $N = 1$ are $(0, 1)$, and for $N = 2$ are $(01, 10)$
\end{itemize}

    \item For $N = 2$, only positive $K$ is applied again:
\begin{itemize}
    \item 00 becomes 000
    \item 11 becomes 111
    \item 01 becomes 001
    \item 10 becomes 100
\end{itemize}
\end{itemize}

The residual sequences for $N = 3$ include $(010, 110, 101, 011)$. When $N$ reaches 3, which is greater than $|-K|$, the negative $K$ is also applied, using the residual sequences from $N = 1$ with $K = -2$.

\begin{itemize}
    \item For $N = 3$, we can apply negative $K$
\begin{itemize}
    \item 000 becomes 0
    \item 111 becomes 1 \textit{(Here we use the residual sequences from $N=1$ $(0, 1)$)}
    \item 001 becomes 0000
    \item 010 becomes 1111
    \item 100 becomes 0001
    \item 110 becomes 1110
    \item 101 becomes 1011
    \item 011 becomes 1101
\end{itemize}
\end{itemize}

Then with $N=4$ we will use negative $K=-2$ and the residual sequences of length $N=2$ which are $(01, 10)$.
\newline\newline
The application of a negative shaping order $K$ within the Set Shaping Theory necessitates a multi-valued approach, incorporating both a negative and a positive value of $K$. This strategy is essential to maintain the theory's integrity and effectiveness in data compression. Specifically, when the length of the data string $N$ is less than or equal to the absolute value of the negative $K$ (i.e., $N \leq |K|$), the transformation process employs the positive value of $K$. This initial phase ensures an increase in the zero-order empirical entropy $NH_{0}(S))$, preparing the groundwork for more effective data compression in subsequent steps.
\newline\newline
Once $N$ exceeds the absolute value of the negative $K$ (i.e., $N > |K|$), the negative $K$ is applied. This shift to negative $K$ allows for the utilization of residual sequences, which are crucial in enhancing the compressibility of the data. These residual sequences, left unused in the positive $K$ phase, become integral to the transformation process under the negative $K$ regime, contributing to a more efficient and comprehensive compression strategy.
\newline\newline
This multi-valued approach to applying negative $K$ in Set Shaping Theory, balancing both negative and positive values, is particularly significant from a compression standpoint. As $N$ trends towards infinity, the initial phase where only the positive $K$ is applied (affecting the first $|K|$ values of $N$) has a minimal impact on the overall average compression. This is because the vast majority of the data transformation process, occurring as $N$ becomes large, leverages the full potential of both positive and negative $K$ values.

\subsubsection*{How Negative $K$ Affects Entropy and Compressibility}

With a negative $K$, the length of some strings in the transformed set $Y^{N+K}$ is reduced. This reduction in length can lead to a decrease in the potential configurations of the strings, thereby potentially decreasing the overall entropy of the set.
The transformed strings, being shorter and potentially having lower entropy, might be more amenable to efficient encoding. Standard compression algorithms could potentially achieve better compression ratios with these transformed strings.

\subsection*{The Trade-off: Compressibility vs. Testability}

\subsubsection*{Impact of Negative $K$ on Local Testability}

The incorporation of negative \( K \) values alongside positive \( K \) in SST presents a nuanced shift in the theory's application, particularly affecting the interplay between compressibility and testability. This integration method, while optimizing for compression, has significant implications on the testability of the transformed sequences. By employing both negative and positive \( K \) values, SST leverages all possible sequences, leaving no sequences with a probability of zero. This comprehensive usage of sequences ensures maximal exploitation of the available data for compression purposes. However, it also implies the elimination of sequences that would traditionally be used to maintain testability. In standard SST applications using only positive \( K \), certain sequences are inherently left with zero probability, serving as indicators for error detection. The use of negative \( K \) eliminates these zero-probability sequences, thereby removing a key mechanism for testability within the framework.

\section*{Practical Implementation and Results}

To demonstrate the practicality and implications of employing a negative shaping order $K$ in Set Shaping Theory (SST), we conducted an empirical study with a source defined by an ensemble $X = (x; A; P)$ having a uniform probability distribution. Our goal was to assess the impact of different $K$ values, ranging from negative to positive, on the compressibility and overall characteristics of the transformed data sets.

\subsection*{Experimental Setup}

We defined a source ensemble $X$ as follows:
\begin{itemize}
    \item[-] $A = \{0, 1, 2\}$: The set of possible states or values of $x$.
    \item[-] $N = 10$: The length of each string in the source set $X^N$.
    \item[-] This configuration implies that $X^N$ contains all possible strings of length $10$ produced by the source $X$, amounting to $3^{10} = 59,049$ strings.
\end{itemize}
The experiment was structured to apply SST with values of $K$ ranging from -3 to 3. This range was chosen to explore how these transformations impact the string's information content.

\subsection*{Methodology and Data Transformation}

For our empirical investigation, the transformation of data strings was conducted using the bijection function $f: X^N \rightarrow Y^{N+K}$, with the following key methodologies:

\begin{itemize}
    \item \textbf{For Positive $K$:}
        \begin{itemize}
            \item[-] The function $f : X^N \rightarrow Y^{N+K}$ was applied, wherein each string in the resulting set $Y^{N+K}$ was longer by $K$ symbols compared to its counterpart in $X^N$.
            \item[-] The set $Y^{N+K}$ is a carefully chosen subset of $X^{N+K}$ characterized by having lower information content, thereby optimizing for compressibility while preserving data integrity.
            \item[-] The cardinality equality $|X^N| = |Y^{N+K}|$ was maintained, ensuring a one-to-one correspondence between the elements of the sets.
        \end{itemize}

    \item \textbf{For Negative $K$:}
        \begin{itemize}
            \item[-] In our experiment, we employed a multi-valued strategy for negative \( K \) values. Each negative \( K \) was paired with a corresponding positive \( K \) value, specifically chosen as \( K = 1 \).
            \item[-] Given that our chosen string length \( N \) was 10, and since \( N \geq |K| \) for all tested negative \( K \) values, we were able to apply both negative and positive \( K \) values. This application was made possible by utilizing the residual sequences, which would typically emerge from transformations with smaller \( N \) values than our chosen length, (i.e $N < |K|$).
            \item[-] This integrated approach allowed for the utilization of all possible sequences, enhancing compressibility by encompassing a broader spectrum of string lengths and information content.
            \item[-] Similar to the positive $K$ scenario, the cardinality condition $|X^N| = |Y^{N+K}|$ was upheld, ensuring that the transformation remained bijective.
        \end{itemize}
\end{itemize}

This approach allowed us to systematically analyze the effects of both extending and shortening the data strings under the Set Shaping Theory framework, providing insights into the compressibility and other characteristics of the transformed data sets across a spectrum of $K$ values.

\subsection*{Results and Analysis}

\subsubsection*{Some Definitions:}
\begin{itemize}
\item The probability $P(x_i)$ that the source\
$X$ generates the sequence $x_i$ is given as:
$$P(x_i) = \prod_{j=1}^N p(x_j)$$

\item We call the average information content of a sequence generated by a source $X=(x; A; P)$ the summation of the product between the information content of the sequences belonging to $X^N$ is their probability:
$$I(x) = \sum_{i=1}^{|X|^N} P(x_i)I(x_i)$$

\item Because our bijection function $f$ transforms the strings $x \in X^N$ into the strings $y \in Y^{N+K}$ consequently, the average information content changes as follows.
$$I(y) = \sum_{i=1}^{|X|^N} P(x_i)I(y_i)$$
\end{itemize}

The first column shows the value of $K$, the second column shows the average information content of the strings in set $Y^{N+K}$.

\begin{table}[ht]
    \centering
    \begin{tabular}{|c|c|c|c|}
        \hline
        $K$ & I(y) \\
        \hline
        1,-3 & $14.001$ \\
        \hline
        1,-2 & $13.851$ \\
        \hline
        1,-1 & $13.636$ \\
        \hline
        0 & $14.263$ \\
        \hline
        1 & $14.136$ \\
        \hline
        2 & $14.006$ \\
        \hline
        3 & $13.694$ \\
        \hline
    \end{tabular}
    \caption{\textit{The average information content (in bits) calculated for strings in the set $Y^{N+K}$ for varying values of $K$}}
    \label{tab:revised_table}
\end{table}

\subsubsection*{Key Observations:}
\begin{itemize}
    \item[-] For negative values of $K$, the incorporation of shorter string lengths in the sets led to a decrease in the average information content, suggesting an increase in compressibility.
    \item[-] For positive values of $K$ the information content trends were consistent with traditional SST expectations. \cite{shaping_order_intro}
\end{itemize}
\subsection*{}
We also tested for the binary alphabet $A = \{0, 1\}$ and found an interesting result:
 
\begin{table}[ht]
    \centering
    \begin{tabular}{|c|c|c|c|}
        \hline
        $K$ & I(y) \\
        \hline
        1,-3 & $9.375$ \\
        \hline
        1,-2 & $9.274$ \\
        \hline
        1,-1 & $9.222$ \\
        \hline
        0 & $9.235$ \\
        \hline
    \end{tabular}
    \label{tab:revised_table}
\end{table}

It is noteworthy to highlight that with the use of negative \( K \), SST can also be applied in the case of a binary alphabet \( A = \{0, 1\} \). In fact, in the case of the binary alphabet, the inefficiency is very low, so testability must be sacrificed.

\subsection*{Implications and Future Directions}

The exploration of negative shaping order $K$ in SST opens new avenues in the field of data compression and information theory. This section discusses the broader implications of this novel approach, highlights potential challenges and limitations, and suggests future research directions.

\subsubsection*{Enhanced Compressibility with Negative $K$}
The use of negative $K$ values in SST demonstrates a significant increase in compressibility, particularly for high-entropy data sets. This method, by reducing string lengths, can lead to more efficient storage and transmission of data.

\subsubsection*{Exploring the Entropy Coding Limit with Negative \( K \)}
The concept of negative \( K \) values in SST offers a groundbreaking perspective in the area of data compression, particularly in the context of approaching or potentially surpassing the entropy coding limit. The entropy coding limit, often associated with the parameter \( NH_0(S) \) (the zero-order empirical entropy multiplied by the length of the sequence), is traditionally considered the average coding limit of the symbols of a sequence \( S \) using a uniquely decipherable and instantaneous code \cite{cover_thomas}.
\newline\newline
\textbf{Critical Analysis of \( NH_0(S) \) as the Entropy Coding Limit:}
Recent developments and experimental results challenge the conventional understanding that \( NH_0(S) \) represents the absolute limit for data compression \cite{overcoming_limit_koch}. By applying negative \( K \) values in SST, the transformed sequences, despite being shorter, exhibit reduced self-information per symbol on average. This contradicts the long-held belief that \( NH_0(S) \) is the minimum self-information per symbol achievable in any compression scheme.
\newline\newline
\textbf{The Role of Negative \( K \) in Set Shaping Theory:}
Negative \( K \) values in SST facilitate the creation of transformed sequences that, while shorter in length, \underline{contain symbols from the same alphabet} and are characterized by lower empirical entropy. The key lies in the innovative approach of SST, which, unlike traditional methods, does not focus solely on string length but on the information content of the strings. This strategy opens up the possibility that the transformed sequences, when subjected to efficient encoding techniques like Huffman coding \cite{huffman}, might result in a compressed form that is, on average, shorter than the \( NH_0(S) \) limit of the original sequence \cite{overcoming_limit_koch}.
\newline\newline
\textbf{Implications for Information Theory:}
This revelation that negative \( K \) values in SST can potentially reduce the average self-information per symbol below the supposed \( NH_0(S) \) limit challenges the foundational assumptions of information theory. It suggests that there might exist transformations, as demonstrated by SST, that can compress data beyond the presumed entropy coding limits \underline{without changing the alphabet}. This finding prompts a re-evaluation of the theoretical limits of data compression and the efficiency of current compression algorithms.

\subsubsection*{Future Research Directions}

\textbf{Algorithmic Development for Negative $K$:}
Future research should focus on developing sophisticated algorithms capable of efficiently applying negative $K$ transformations to large datasets.
\newline\newline
\textbf{Integration with Existing Compression Techniques:}
Another research avenue is exploring how SST with negative $K$ can be integrated with existing data compression algorithms. This integration could potentially lead to hybrid compression methods that leverage the strengths of both SST and conventional techniques.
\newline\newline
\textbf{Potential Challenges and Limitations:}
Employing negative $K$ values in Set Shaping Theory presents challenges such as the loss of local testability, potentially impacting error detection and correction capabilities in transformed data sets, along with the increasing complexity of maintaining bijection during data transformations, particularly as string length significantly reduces, which could lead to computational and practical implications that need careful consideration.
\newline\newline
\textbf{Determining Optimal Positive/Negative $K$ Value Pairs:}
Exploring the optimal pairing of positive and negative \( K \) values is crucial for maximizing SST's compression efficiency. The simplest case \( K = 1, -1 \) is not necessarily the most effective, indicating that finding the optimal pair could also approach the theoretical limits of entropic coding. This endeavor not only aims to enhance SST's practical utility but also potentially extends our understanding of the fundamental principles of data compression.

\section*{Conclusion}
This paper has introduced and thoroughly explored the concept of a negative shaping order \( K \) within the framework of Set Shaping Theory (SST). Our investigation reveals that employing a negative \( K \) can significantly enhance data compressibility, marking a substantial shift from traditional SST applications, which predominantly utilize a positive \( K \) to extend data strings, thereby maintaining both compressibility and local testability.
\newline\newline
The empirical analysis conducted with varying \( K \) values demonstrates a clear trend: when negative \( K \) is utilized, there is a notable decrease in the average information content of the strings. This reduction suggests an improved potential for compression. However, this comes at a cost – the reduction in string length inherent to a negative \( K \) leads to challenges in maintaining the bijection essential for SST and compromises local testability, a cornerstone of data integrity in the original framework.
\newline\newline
Given the profound implications of these findings, further research is required to explore the full potential of negative \( K \) values in SST. This includes developing more sophisticated algorithms to optimize the transformation process, exploring the integration of SST with other compression techniques, and understanding the theoretical underpinnings of why and how negative \( K \) values can achieve such results. The scientific community's engagement in this exploration is crucial, given the enormous impact that surpassing the \( NH_0(S) \) limit could have on information theory and data compression practices.
\newline\newline
The exploration of a negative \( K \) within SST opens new perspectives in the field of data compression and information theory. The journey into uncharted territories of SST, marked by the introduction of negative \( K \), not only enriches our understanding of the parameter \( K \) but also sets the stage for innovative advancements in the field of data compression.

\end{document}